# Multiple damage detection in piezoelectric ceramic sensor using point contact excitation and detection method

Sayantani Bhattacharya[1], Nitin Yadav[2], Azeem Ahmad[3], Frank Melandsø[3], and Anowarul Habib[3] *

[1] Dept. of Mining Machinery Eng., Indian Institute of Technology Dhanbad. India
[2] Dept. of Electrical Eng., Indian Institute of Technology Delhi, India
[3] Dept. of Phys. and Tech, UiT-The Arctic University of Norway, Tromsø, Norway

*e-mail: anowarul.habib@uit.no

**Abstract:**
Lead Zirconate Titanate [$(Zr_xTi_{1-x})O_3$] is used to make ultrasound transducers, sensors, and actuators due to its large piezoelectric coefficient. Several surface and subsurface micro defects develop within the Lead Zirconate Titanate (PZT) sensor due to delamination, corrosion, huge temperature fluctuation, etc., causing a decline in its performance. It is thus necessary to identify, locate, and quantify the defects. Non-Destructive Structural Health Monitoring (SHM) is the most optimal and economical method of evaluation. Traditional ultrasound SHM techniques have a huge impedance mismatch between air and any solid material. And most of the popular signal processing methods define time-series signals in only one domain which gives sub-optimal result. Thus to improve the accuracy of detection point contact excitation and detection method have been implemented to determine the interaction of ultrasonic waves with microscale defects in the PZT. And Haar Discrete Wavelet Transformation (DWT) is applied upon the time series data obtained from coulomb coupling setup. Using the above process, defect up to 100 μm diameter could be successfully distinguished and localised.

## 1. Introduction:

Lead zirconate titanate $(Zr_xTi_{1-x})O_3$ (PZT) is a piezoelectric material with a large piezoelectric coefficient. PZT ceramics have negligible mass, have easy and fast integration, large frequency responses, low power consumption, and low cost of the sensors, superior electromechanical coupling, and impedance matching with various substrates [1-3]. These properties make PZT ceramics extremely suitable for integration into the host structure as an *in-situ* generator/sensor, and thus used extensively.

In the harsh and extreme environmental conditions, such as corrosion, fatigue, and delaminating due to extreme temperature fluctuations, several surface and subsurface micro defects are likely to be introduced within the sensor. These flaws may also be intrinsic to the bulk material as seldom they are introduced in the final stages of the fabrication or early stages of device operation. It is thus necessary to identify, locate, and quantify the defect in sensors to avoid structural failure and false alarm in structural health monitoring (SHM) applications.

In the past several years, a wide range of innovative ways have been implemented for NDE techniques for detecting intrinsic and bulk damage of ceramic components [4-8]. The air-coupled ultrasonic techniques are being increasingly used for material characterization, non-destructive evaluation of composite materials using guided waves as well as for distance measurements [9,10]. The main drawback of air-coupled ultrasound is the huge impedance mismatch between air and any solid material. Apart from ultrasonic methods, several optical methods have been considered to identify the surface and subsurface flaws in PZT ceramics. The most common optical measurements include photoacoustic microscopy, optical coherence tomography, and optical gating technique [11-13]. The scanning laser Doppler vibrometer (SLDV) has been employed for three dimensional visualization of acoustic waves interference with inclusions and flaws in metallic structures, piezo-electric crystals, and piezo ceramics [14,15]. Other, notable signal processing techniques that are widely accepted in SHM applications are, Singular Spectrum Analysis, Frequency domain decomposition (FDD), Auto-Regressive Model, and Extended Kalman Filter Weighted Global Iteration techniques [16-22]. These techniques provide a damage index parameter based on the spectral content or statistical evaluation of the time series. Generally, processing the stationary signal is easier as with help of entire data and the statistics of signal can be evaluated and this information can be used to derive conclusion. Whereas, for non-stationary signals to derive any information, algorithms that are adaptive to use. Bodeeux and Golinval [23], Sohn et. al. [24,25], and Yao and Pakzad [26] have developed adaptive feature extraction approach. These approaches assume behavior of structure to be linear and detect the damage with help of changes in extracted feature or the proposed novel damage index derived from these features [27].

In last several years, a significant amount of effort is devoted for improving the point contact excitation and detection method to excite the acoustic waves in piezoelectric crystals and ceramics [28-34]. The point contact excitation and detection method is a unique way to generate acoustic waves in piezoelectric materials in the absence of coupling media, mechanical, geometrical and electrical resonances, and photolithography. The working principle of this technique depends on the transfer of electromagnetic field to mechanical energy to excite phonon vibration in piezoelectric materials [42]. The Coulomb coupling method and spectral decomposition technique has been implemented for localization of surface defect in piezo-ceramic structures wherein the signal processing is done using Fast Fourier Transform [42]. Unlike Fast Fourier Transform, Wavelet transformation has the ability to define any type of signal in both time and frequency domain simultaneously and has fast computation. Thus, to improve the accuracy further, this paper aims to implement the Wavelet transformation technique of signal processing to localize the defects in piezo-ceramic sensors using coulomb coupling method.

## 2. Experimental Setup:

Our group has previously provided a complete overview of the experimental setup, working principle as well as the excitation and detection probe fabrication [28-36]. The complete experimental setup consists of 4 basic processes i.e., 1) probe fabrication for both sending and receiving electrodes, 2) PZT sample preparation, 3) damage insertion on the PZT sample, and 4) data acquisition.

The damage detection and degradation of the sample was demonstrated on a 3 mm thick PZT ceramics with a dimension of 20 ×20 mm$^2$. The PZT ceramic was chemically etched using Ferric Chloride solution in order to remove the silver (Ag) electrodes from both sides of the sample. The sample was then cleaned with acetone and distilled water and later on, dried with nitrogen. The PZT sample was placed on the top of receiving electrode in the middle of the scan area. The data collection started with a healthy state scan with a scan area of 10 ×10 mm$^2$, performed on the surface of the PZT ceramic sample.

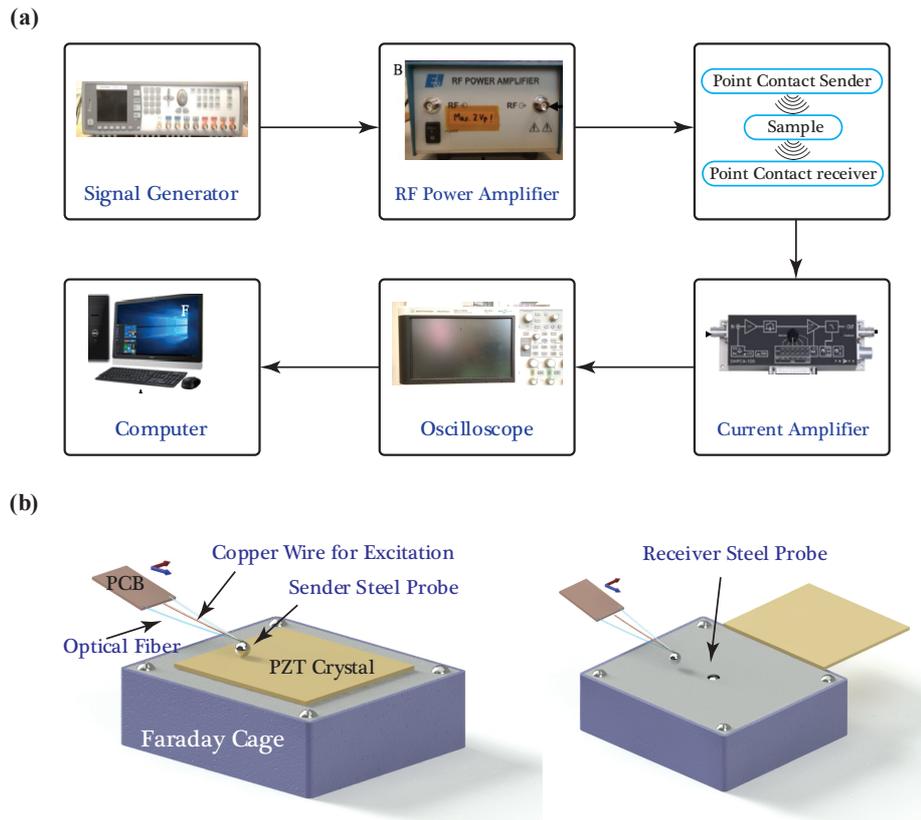

**Figure 1:** (a) Experimental setup for point contact excitation and detection scheme and (b) 3D illustration of the arrangement of sender and receiver steel probe along with the PZT sample.

The sample was then introduced with a circular hole of diameter 500 μm using a high-speed drill. The depth of the holes was fixed to 2.8 mm (approximately) for all the damages. After, the 1st damage insertion a line scan was performed in X direction and data was collected from 12 different positions with a step size of 82 μm. Later on, the size of the hole was enlarged sequentially by employing 600 μm, 800 μm, and 900 μm drill bit and data was collected for each damage dimension. After 900 μm, another 500 μm hole was drilled next to the previous hole, and this damage insertion process continued four times. Thus the final sample had 4 holes of 900 μm diameter. The stage where PZT sample had 1 hole is referred to as Damage state 1, when sample had 2 holes is referred to as Damage state 2 and so on.

## 3. Methodology:

The major challenge in structural health monitoring is the detection of the micro damages in the solid structure. There are various techniques mentioned above by which one can detect the damages i.e., combination of various signal processing and analytical techniques. This paper focuses on point contact excitation and detection method along with two different signal processing techniques for damage detection: i) analysis of the power spectral density of the signals, and ii) analyzing coefficients by Haar (db1) discrete wavelet transformation (DWT) of the signals.

### 3.1 Power Spectral Density analysis of the signal

Fourier Transform decomposes time-series data into sum of infinite sine and cosine functions of different amplitudes and frequencies. This process converts a waveform that is difficult to describe mathematically, into more manageable series of trigonometric functions. Which when added together exactly reproduces the original waveform. Fourier Transforms are of two types Discrete and Continuous. Wherein distinct ordered pairs representing the original input function are equally spaced in their input variable (equal time steps, in this case) are called Discrete Fourier transform (DFT) while ordered pair with input variables within infinitesimal difference between them are called Continuous Wavelet Transform.

Fourier Transform (F) of a function f (t) is given by the following expression:

$$F(w) = \int_{-\infty}^{\infty} f(t) e^{-2\pi i t w} dt \quad (1)$$

Where, t is time, w is frequency in Hertz.

Discrete Fourier Transform (S) of a function f(x) is given by the following expression:

$$S(w) = \sum_{i=0}^{N-1} e^{-jw(i\Delta t)} y_i(i\Delta t) \Delta t \quad (2)$$

Where, N is total no. of equally spaced data points and $y_i(i\Delta t)$ is the actual data recorded at $i^{th}$ time.

Many time-series functions, show complicated periodic behavior. Spectral analysis is a technique that helps in discerning these underlying periodicities. To perform spectral analysis, data is first transformed using Fast Fourier Transform (FFT) from time domain to frequency domain. FFT is a computationally efficient algorithm for solving DFT faster, by reducing the number of redundant calculations. Power Spectral Density of a signal analyses the distribution of power as a function of frequency, over the entire frequency range.

Mathematical representation of PSD is:

$$P_{ab} = \frac{1}{2\pi} \int_{wa}^{wb} S(w) \, dw \quad (3)$$

### 3.2 De-noising by Wavelet transformation (DWT) of the signal

In Method 1, normalized FFT is used. Unlike FFT, Wavelet Transform (WT) can extract information from both spectral and temporal regions thus ensuring more resolute signal processing. WT decomposes a signal into multiple lower resolution levels by varying the scaling and shifting factors of a single wavelet function (mother wavelet). In the first step, the time-series is decomposed into two high and low frequency components. Then, high frequencies are retained, while low frequencies are decomposed again into two high and low frequencies. High frequencies are called details coefficient and low frequencies are approximation coefficient of the signal. Wavelet transforms are of two types, Continuous wavelet transform (CWT) and Discrete wavelet transform (DWT). CWT is very slow due to the extra data that overlaps with its neighboring data (duplicity). Therefore, DWT is used in this paper. This paper uses the first member of the Daubechies family of orthogonal discrete wavelet, popularly known as the Haar Wavelet. One of the main reasons behind choosing Haar wavelet is that it is computationally fast and memory efficient, as it can be calculated in place without the need of temporary array allocation. Haar wavelet is a discontinuous step function. These abrupt changes in the function are beneficial for the analysis of signals with sudden transitions. The high-pass (G) and low-pass (H) filters is given by $G = \begin{bmatrix} 1/\sqrt{2} & 1/\sqrt{2} \end{bmatrix}$ and $H = \begin{bmatrix} 1/\sqrt{2} & 1/\sqrt{2} \end{bmatrix}$.

$$\text{Mother Wavelet of Haar }((n)) = \begin{cases} 1 & 0 \leq n \leq 1/2 \\ -1 & \frac{1}{2} \leq n \leq 1 \\ 0 & otherwise \end{cases}$$

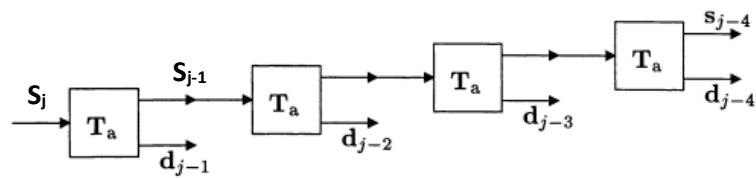

**Flowchart 1:** The above flowchart represents DWT over 4 scales, wherein j is the number of times the transformation algorithm is applied to the original signal $(S_j)$.

Where,

$S_j$ – Original input signal

$S_{j-1}$ – Approximation coefficient at level 2

$d_{j-1}$ – Details coefficient at level 2

$T_a$ - Direct transform

$T_a$ is building block of DWT. The formulae used in a $T_a$ block of Haar Transform:

1. $d_{j-1}[n] = S_j[2n + 1] - S_j[2n]$
2. $S_{j-1}[n] = S_j[2n] - \frac{1}{2} * d_{j-1}[n]$
3. $S_{j-1}[n] = \sqrt{2S_{j-1}[n]}$
4. $d_{j-1}[n] = 1 \big/ \sqrt{2d_{j-1}[n]}$

## 4. Procedure:

### 4.1 Power spectral analysis on received signal

A steel probe receives signal in the form of time-series data. For the same damage state and size, 12 different data sets are measured from uniformly distributed axis points. To extract

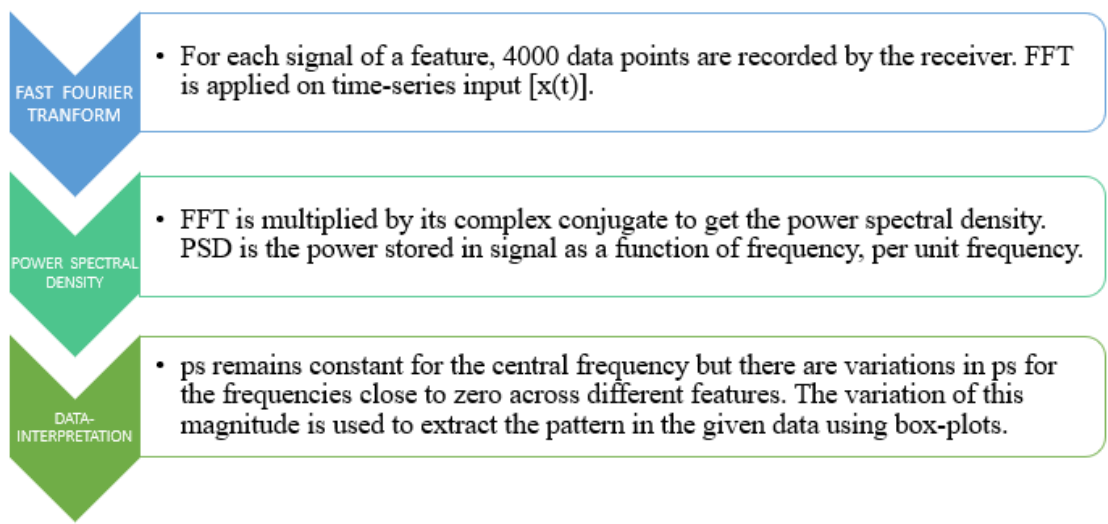

- **FAST FOURIER TRANFORM**: For each signal of a feature, 4000 data points are recorded by the receiver. FFT is applied on time-series input [x(t)].
- **POWER SPECTRAL DENSITY**: FFT is multiplied by its complex conjugate to get the power spectral density. PSD is the power stored in signal as a function of frequency, per unit frequency.
- **DATA-INTERPRETATION**: ps remains constant for the central frequency but there are variations in ps for the frequencies close to zero across different features. The variation of this magnitude is used to extract the pattern in the given data using box-plots.

meaningful information power spectral (ps) analysis is performed. ps value is calculated for each time-series of 4000 data points, and 12 such values were obtained. This was then converted into a box plot for that damage state. Boxplots of four damage states are then compared in Figure 2.

**Flowchart 2:** The flowchart explains the three steps (i.e. Fast Fourier Transformation, power spectral density calculation and data interpretation) performed to obtain the box-plots from the experimental data.

### 4.2 Haar wavelet analysis on received signal

A steel probe receives signal in form of time-series data. For the same damage state and size, 12 different data sets are measured from uniformly distributed axis points. To extract meaningful information from the data, De-noising is performed using Haar discrete wavelet analysis. The final output is the summation of detailed and approximation coefficients of all the six levels. Approximation coefficients are equated to zero and average of the values thus obtained are considered 12 times to develop the box plot of a particular damage size and damage state. For each damage state, four damage sized box plots are compared and in total four such damage states are used in this experiment.

**Flowchart 3:** The above flowchart explains the four steps (i.e. Zero-Padding, DWT Calculation, de-noising and data interpretation) performed to obtain the box-plots from the experimental data.

## 5. Experimental Observation:
### 5.1 Power Spectral Density analysis

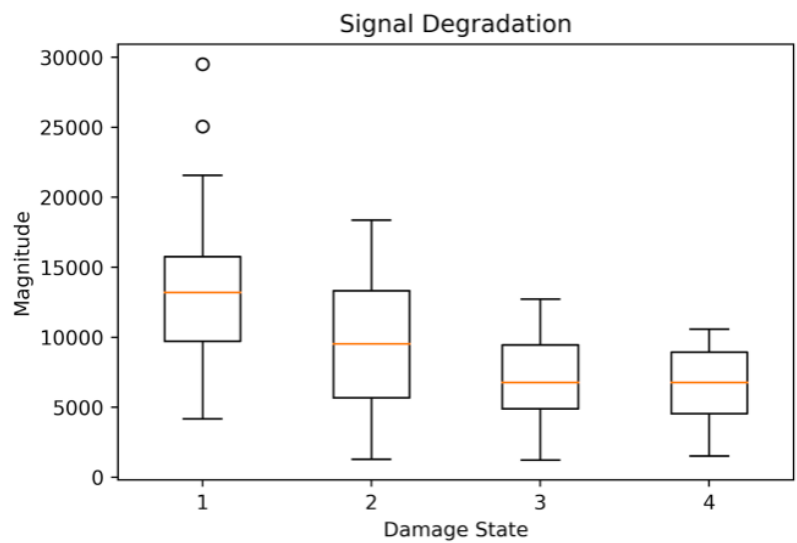

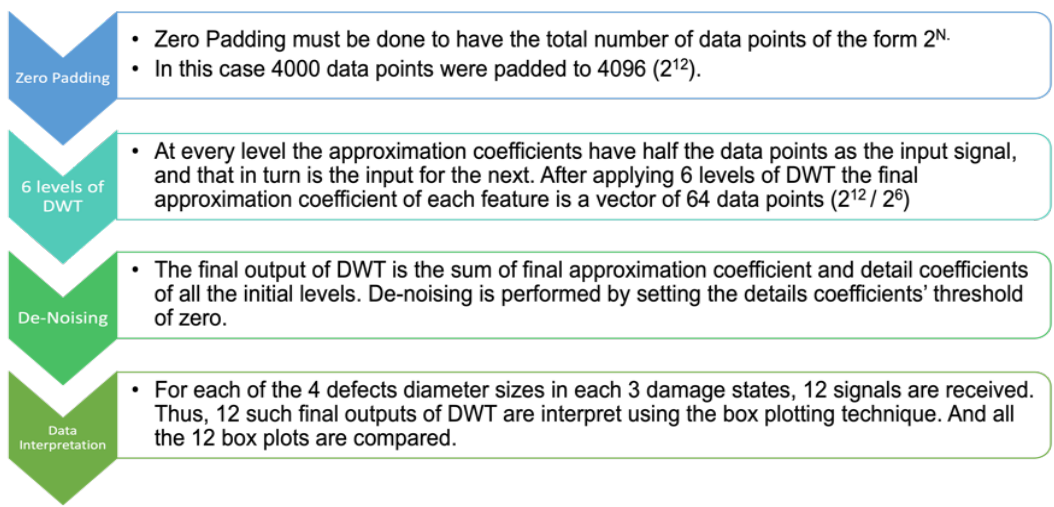

**Figure 2:** The individual boxplot computes the variation of second maxima amplitude value of PSD plots of the time-series function of the 12 features. Thus, it approximately represents the power stored by those signals. The comparison between four such damage states is shown above for damage size of 500μm.

It can be observed from Figure 3 that with the increase in damage state, the statistical median of power reduces. Observations of the figures shown above are as per the expectations because with the increase in damage in the sample, the signal will get interrupted to a greater extent leading to loss of information and reduction of power. Similar trend is observed for damage sizes of 600, 800 and 900 μm. This variation in median values is distinct enough to distinguish between the different damage states. Thus, it can be concluded that PSD analysis method can be effectively used for understanding the degree of damage in PZT ceramics.

## 5.2 Discrete Wavelet transformation

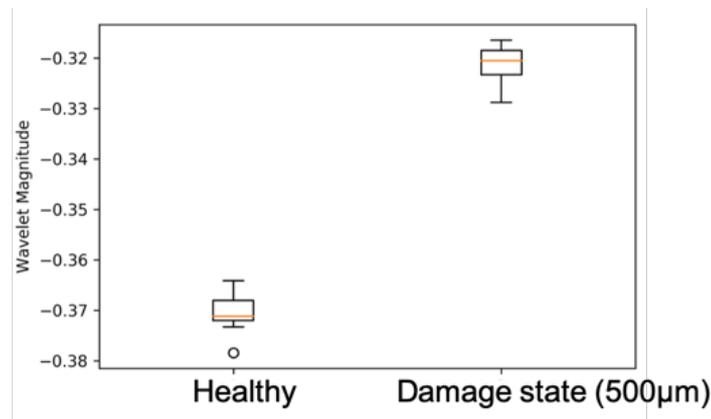

**Figure 3:** The above boxplots are of the de-noised signal, as described in procedure. The plot in the left is of a healthy PZT sample, and the one in right is of a PZT ceramic sample with 500μm damage size in 1st damage state.

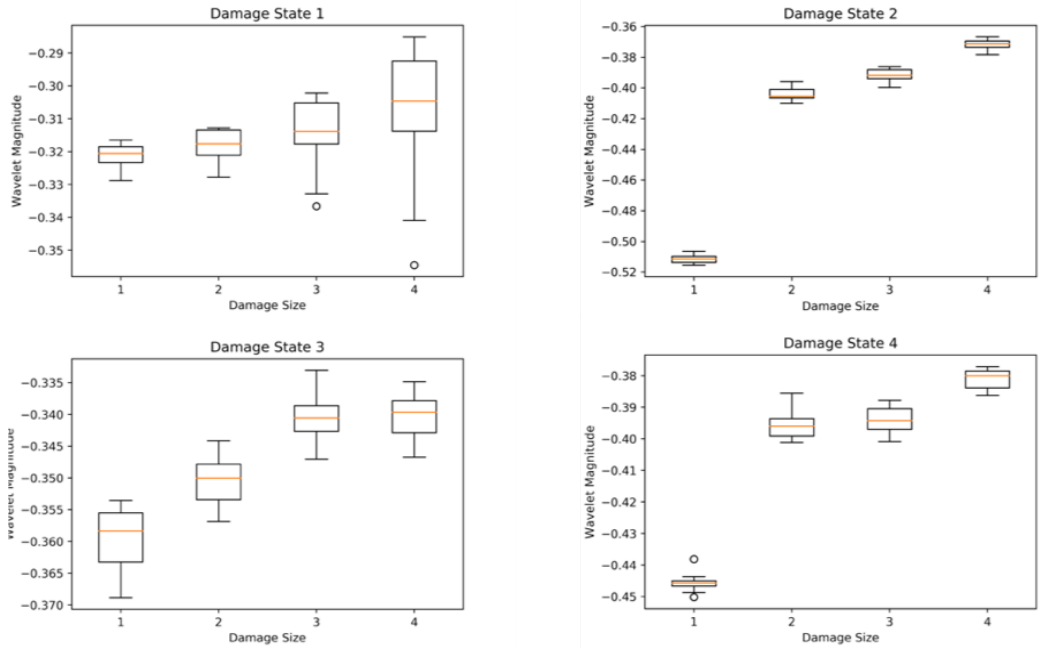

**Figure 4:** Damage size 1, 2, 3, and 4 corresponds to damage sizes 500, 600, 800 and 900 μm respectively. Each quarter of the above plots shows comparison between the different damage sizes of a particular damage state. Boxplot of all the 4 damage states is shown above, in ascending order of progression

A clear distinction can be observed in the statistical medians of the box plots in figures 4 and 5. Figure 4 proves that Haar DWT De-noising method efficient for damage detection. And it can be concluded from Figure 5 that DWT is more efficient compared to PSD analysis. PSD was successful in distinguishing between the damage states of PZT sample, while DWT can also distinguish between different damage sizes (with 100 μm variation in dimension) within the same damage state.

# 6. Conclusion

This paper targets a common but challenging problem of damage detection and damage classification for small differences in the damage size. Point contact excitation and detection method is used in this experiment to extract raw data. As the signals are very sensitive to noises, it is very difficult to interpret anything from the raw data. Therefore, a couple of signal processing techniques (i.e. Power spectral analysis and Wavelet transformation) are used to extract meaningful information from the main signal. In power spectral analysis, the phenomenon that structural damage leads to power reduction, is used to discern different damage states. And Discrete wavelet transformation having more resolution is used to distinguish between the damage states as well as between damage sizes up to difference of only 100μm, for the same damage state. Thus, according to the results, the proposed methods were successful to distinguish PZT ceramic samples based on structural damage severity.

## 7. References

1) V. Rathod and D.R. Mahapatra: NDT & E Int. 44 [7] (2011) 628.

2) H. Mei and V. Giurgiutiu: Struct. Health Monit. (2018) 690.

3) L. Su, L. Zou, C.-C. Fong, W.-L. Wong, F. Wei, K.-Y. Wong, R. S.S.Wu and MengsuYang: Biosens. Bioelectron. 46 (2013)155.

4) S. Quek, P. Tua and J. Jin: J. Intell. Mater. Syst. Struct. 18 [9] (2007) 949.

5) C.R. Farrar and K. Worden: Philos. Trans. Royal Soc. A. 365 [1851] (2007) 303.

6) C. Boller, C. Biemans, W.J. Staszewski, K. Worden and G.R. Tomlinson: Smart Structures and Materials 1999: Smart Structures and Integrated Systems, 1999, p. 285.

7) M. Todd, J. Nichols, L. Pecora and L. Virgin: Smart Mater. Struct. 10 [5] (2001)1000.

8) J.-B. Ihn and F.-K. Chang: Struct. Health Monit. 7 [1] (2008) 5.

9) R. Kazys, R. Sliteris and J. Sestoke: Sensors. 17 [1] (2017) 95.

10) Z. Fan, W. Jiang and W.M. Wright: Ultrasonics. 89 (2018) 74.

11) R. Komanduri, J. Lange, J.P. Wicksted and J.S. Krasinski: ARPA Ceramic Bearing Technology Annual Review. (1994)10.

12) M. Bashkansky, M. Duncan, M. Kahn, D. Lewis and J. Reintjes: Opt. Lett. 22 [1] (1997) 61.

13) P.R. Battle, M. Bashkansky, R. Mahon and J.F. Reintjes: (Springer, Boston, 1996) pp. 1119.

14) T.E. Michaels, J.E. Michaels and M. Ruzzene: Ultrasonics. 51 [4] (2011) 452.

15) Z. Tian and L. Yu: J. Intell. Mater. Syst. Struct. 25 [9] (2014) 1107.

16) N.E. Huang, Z. Shen, S.R. Long, M.C. Wu, H.H. Shih, Q. Zheng, N.-C. Yen, C.C. Tung and H.H. Liu: Proc. Royal Soc. A, 1998, p. 903.

17) N.E. Huang, M.-L.C. Wu, S.R. Long, S.S. Shen, W. Qu, P. Gloersen and K.L. Fan: Proc. Royal Soc. A, 2003, p. 2317.

18) M.E. Wall, A. Rechtsteiner and L.M. Rocha: A practical approach to microarray data analysis (Springer, 2003) p. 91. AUTHOR SUBMITTED MANUSCRIPT - JJAP-S1102590.

19) J. Shlens, A tutorial on Principal Component Analysis: Derivation, Discussion and Singular Value Decomposition (2003)

20) C.R. Farrar and K. Worden: Philos. Trans. R. Soc. London, Ser. A. 365 [1851] (2006) 303.

21) L. Jeen-Shang and Z. Yigong: Comput. struct. 52 [4] (1994) 757.

22) H. Sohn and C.R. Farrar: Smart Mater. Struct. 10 [3] (2001) 446.

23) J.-B. Bodeux and J.-C. Golinval: Smart Mater. Struct. 10 [3] (2001) 479.

24) H. Sohn, J.A. Czarnecki and C.R. Farrar: J. Struct. Eng. 126 [11] (2000) 1356.

25) H. Sohn, C.R. Farrar, N.F. Hunter and K. Worden: J. Dyn. Syst. Meas. Contr. 123 [4] (2001) 706.

26) R. Yao and S.N. Pakzad: Mech. Syst. Sig. Process. 31 (2012)355.

27) K. Worden, C.R. Farrar, J. Haywood and M. Todd: Struct. Control Health. Monit. 15 [4] (2008)540.

28) A. Habib, E. Twerdowski, M. von Buttlar, M. Pluta, M. Schmachtl, R. Wannemacher and W. Grill: Proc. SPIE, 2006, p. 61771A.

29) A. Habib, E. Twerdowski, M. von Buttlar, R. Wannemacher and W. Grill: Proc. of SPIE, 2007, p. 653214. 30) A. Habib, A. Shelke, M. Pluta, T. Kundu, U. Pietsch and W. Grill: Jpn. J. Appl. Phys. 51 [7S] (2012)07GB05. 31) A. Habib, A. Shelke, U. Pietsch, T. Kundu and W. Grill: Proc. SPIE, 2012, p. 834816.

32) A. Habib, U. Amjad, M. Pluta, U. Pietsch and W. Grill: Proc. SPIE, 2010, p. 76501T.

33) A. Shelke, A. Habib, U. Amjad, M. Pluta, T. Kundu, U. Pietsch and W. Grill: Proc. SPIE, 2011, p. 798415. 34) A. Habib, A. Shelke, M. Pluta, U. Pietsch, T. Kundu and W. Grill: AIP Conference Proceedings, 2012, p. 247.

35) Sayantani Bhattacharya, Prakhar Kumar, Nitin Yadav, Azeem Ahmad, Frank Melandsø, and A. Habib: Ultrasonics Electronics 2021, 2021, p. 99

36) V. Agarwal, A. Shelke, B. S. Ahluwalia, F. Melandsø, T. Kundu and A. Habib: Ultrsonics Electronics 2018, 2018, p. 2P2-13